\documentclass[aps,floatfix,preprint,nofootinbib]{revtex4}
\pdfoutput=1
\usepackage{graphics}
\usepackage{graphicx}
\usepackage{amssymb}
\usepackage{amsmath}
\usepackage{verbatim}
\usepackage{pstricks}
\usepackage{slashed}

\newcommand{\nc}{\newcommand}
\nc{\beq}{\begin{equation}}
\nc{\eeq}{\end{equation}}
\nc{\barray}{\begin{eqnarray}}
\nc{\earray}{\end{eqnarray}}
\nc{\barrayn}{\begin{eqnarray*}}
\nc{\earrayn}{\end{eqnarray*}}
\nc{\bcenter}{\begin{center}}
\nc{\ecenter}{\end{center}}
\nc{\ket}[1]{| #1 \rangle}
\nc{\bra}[1]{\langle #1 |}
\nc{\0}{\ket{0}}
\nc{\mc}{\mathcal}
\nc{\er}[1]{(\ref{eq:#1})}
\nc{\onehalf}{\frac{1}{2}}
\nc{\partialbar}{\bar{\partial}}
\nc{\psit}{\widetilde{\psi}}
\nc{\Tr}{\mbox{Tr}}
\nc{\ev}{\;\mathrm{eV}}
\nc{\mev}{\;\mathrm{MeV}}
\nc{\gev}{\;\mathrm{GeV}}

\def\chii0{\chi_i^0}
\def\chij0{\chi_j^0}

\newcommand{\gsim}{\lower.7ex\hbox{$\;\stackrel{\textstyle>}{\sim}\;$}}
\newcommand{\lsim}{\lower.7ex\hbox{$\;\stackrel{\textstyle<}{\sim}\;$}}

\begin{document}

\setlength{\baselineskip}{0.22in}

\begin{flushright}MCTP-10-27 \\
\end{flushright}
\vspace{0.2cm}

\title{Dark Moments and the DAMA-CoGeNT Puzzle}

\author{
A. Liam Fitzpatrick$^a$  and
Kathryn M. Zurek$^{b}$
}

\vspace*{0.2cm}

\affiliation{
$^a$Department of Physics, Boston University, Boston, MA  \\
$^b$Michigan Center for Theoretical Physics, Department of Physics, University of Michigan, Ann Arbor, MI 48109
}

\date{\today}

\begin{abstract}
\noindent

We consider the velocity dependence arising from scattering through dark multipole moments, and its effects on the consistency of the signals observed by DAMA and CoGeNT with the dark matter hypothesis.
We focus on the effects of the experimental uncertainties on the fits, and show that the two experiments combined favor dark matter scattering with a velocity-dependent cross-section over standard velocity and spin-independent scattering.  When appropriate uncertainties are taken into account, we show that agreement of the two signals with each other and with the results of null experiments can be obtained.

\end{abstract}

\maketitle

\section{The DAMA-CoGeNT Puzzle}

Recently, CoGeNT \cite{cogent} has reported an excess of events at low nuclear recoil.  It is unknown what the source of the events is, but it is consistent with a light Dark Matter (DM) particle in the 7-11 GeV range interacting with their germanium crystals.  The region is remarkably close to the mass and cross-section preferred by the DAMA experiment's \cite{DAMA} observation of an annual modulation \cite{petriello,Chang,Fairbairn,savage}.  The DM mass window in the several GeV range is well-motivated by several models of DM \cite{LDM,mohapatra}, notably models that solve the baryon-DM coincidence problem \cite{ADM}.

At the same time, the preferred cross-section and DM masses of the two experiments, while being close to each other in parameter space, are naively inconsistent with each other.  They are also naively inconsistent with the results of other null experiments, the most relevant of which are the silicon run of CDMS \cite{CDMSIISi,CDMSIISi2} and XENON10 \cite{XENONinelastic}.  The CoGeNT region overlaps neither with the DAMA region for channeled scattering off iodine, nor with the DAMA region for unchanneled scattering off sodium \cite{Fitzpatrick,cogentinterpretations}.  

However, we have also begun to learn more recently about the experimental uncertainties which must be properly taken into account to make a conclusion about the relevance of experimental constraints for excluding the light DM window, as well as for the consistency of the DAMA and CoGeNT signals with each other.  For example, a new measurement of the scintillation light yield efficiency, ${\cal L}_{\rm eff}$ \cite{Manzur}, opens an allowed region for DAMA channeled spin-independent scattering off iodine \cite{Fitzpatrick}.  In addition, by choosing the sodium and germanium quenching factors appropriately, the DAMA unchanneled and CoGeNT spin-independent scattering regions can be marginally consistent with each other \cite{HooperCollar}.  Lastly, by assuming a 20\% systematic uncertainty in the energy threshold for the constraint from CDMS silicon, the region in which the two signals are consistent with each other can be made consistent with the null observation of CDMS.

While obtaining consistency in this way is marginally possible for spin-independent scattering, it does stretch the experimental parameters and theoretical uncertainties to their limits in order to allow it.  The purpose of this paper is two-fold.  First, we unpack the results of \cite{HooperCollar} to show more explicitly how the experimental and theoretical uncertainties allow for better agreement between the signals of DAMA and CoGeNT with each other and with the results of null experiments for standard spin-independent scattering.  We find, however, that pushing all the experimental systematic uncertainties beyond the edge of their $1\sigma$ preferred values only allows agreement between DAMA and CoGeNT in the 99\% C.L. regions.  We then demonstrate that with the choice of a different operator to mediate the scattering, significantly better agreement can be obtained.  Tension with the null results of CDMS-Si and XENON can also be alleviated.

Interestingly, the operators that we consider, the dark anapole and magnetic dipole moment operators,
\begin{eqnarray}
{\cal O}_a &  = & \bar{\chi} \gamma^\mu \gamma_5 \chi A_\mu
\label{anapole} \\
{\cal O}_d & = & \bar{\chi}\sigma^{\mu \nu} \chi F_{\mu \nu}/\Lambda ,
\label{dipole}
\end{eqnarray}
are unique in that the contributions from spin-dependent and spin-independent  scattering  can be equal for some elements (sodium in particular).\footnote{The operator which is usually called the anapole couples to the current, ${\cal O}_a = \bar{\chi} \gamma^\mu \gamma_5 \chi \partial_\nu F_{\mu \nu}$, as discussed in \cite{pospelov}.  This operator has the same spin structure as Eq.~(\ref{anapole}), but has an additional $q^2$ suppression.}
The model that we have in mind is a massive dark photon kinetically mixed with the visible photon.  That the coupling to nuclei in the scattering goes through the SM photon imposes constraints on the coefficients of the scattering cross-section which we utilize.

These operators also have unusual velocity and momentum dependence:
\begin{eqnarray}
\sigma_a &=&   \frac{\mu_N^2 }{4 \pi (q^2+M^2)^2} \left(\left(4 v^2 - q^2\frac{(m_N+m_\chi)^2}{m_N^2 m_\chi^2} \right) F_1^2 + (F_1+F_2)^2 q^2 \frac{2}{m_N^2}\right),
\label{anapolexsec} \\
\sigma_d &=&  \frac{4 \mu_N^2 q^2}{\pi\Lambda^2 (q^2+M^2)^2} \left(\left(4 v^2 - q^2\left(\frac{1}{m_N^2}+\frac{2}{m_N m_\chi}\right) \right) F_1^2 +  (F_1+F_2)^2 q^2 \frac{2}{m_N^2}\right),
\label{dipolexsec}
\end{eqnarray}
with $M$ the mediator mass, $\Lambda$ an expansion parameter (associated in some models with strong coupling in the DM state), $m_N$ the nucleus mass, $m_\chi$ the DM mass, and $q$ and $v$ the momentum transfer and velocity of the incoming WIMP.  We use the standard notation for the form factors $F_1,F_2$ for a coupling of a gauge field to $N$, {\em e.g.}
$
{\cal O}_N \propto i A^\mu \bar{N}\left(F_1 \gamma_\mu + \frac{i F_2}{2 m_N} \sigma_{\mu \nu} q^\nu \right) N
$
when $N$ is spin-1/2.
This unusual momentum and velocity dependence has been noted before in other contexts \cite{cogentinterpretations,pospelov,Sigurdson,Masso,MDDM,bagnasco,ffdm,ffdmchanneled}, though in most of these cases only some of the terms in the full expression are considered (but see \cite{mohapatra}).  We find, by contrast, that both terms arising from the magnetic and electric form factors can be important and give rise to significantly modified spectra.

In this paper we show that non-standard velocity and momentum dependence can, depending on how they enter into scattering cross-section, reconcile the DAMA and CoGeNT regions.
The dark magnetic dipole moment interaction in particular has the right
structure to give agreement between the two experiments, consistent with
null results of other direct detection experiments.  The dark anapole
interaction on the other hand does not bring the two experimental regions
together, and its main benefit is to alleviate tension between DAMA and
the null results.     The magnitude of the shifts in the preferred DAMA and CoGeNT regions, and whether this leads to better agreement, is a detailed numerical question.  This can however be understood qualitatively as follows.
CoGeNT records slightly lower momentum transfer than DAMA, and since these operators are momentum suppressed, this causes CoGeNT to shift slightly up relative to DAMA in comparison to the standard spin-independent case.  More importantly for these operators, however, is the velocity dependence.  The maximum momentum transfer depositable in the detector is $2 \mu_N v$, where $\mu_N$ is the DM-nucleus reduced mass.  Since the typical momentum transfers observed by CoGeNT are lower than those observed by DAMA, and the reduced mass of germanium is higher than that of sodium, the typical velocities of particles observed by CoGeNT are significantly lower than those of DAMA, further suppressing the overall scattering rate in CoGeNT, and further shifting the CoGeNT region up relative to DAMA.  The relative importance of these effects and the amount that it improves the agreement between DAMA and CoGeNT is a question we address in detail in this paper.  

The types of models in which the anapole and magnetic moment operators dominate the scattering are not hard to construct.  For a Majorana particle scattering through a vector mediator, for example, ${\cal O} = \bar{\chi} \gamma^\mu  \chi \bar{N} \gamma_\mu N$ vanishes, and one expects the anapole to be the dominant contribution.  For Dirac particles, one can explicitly construct models where the coupling of the DM particles is purely axial.   Likewise, the DM can have a large magnetic moment when constituents charged under a dark force are bound into a neutral state.  Because the rates are velocity and momentum suppressed, the corresponding scattering cross-section must be large.  This can be accommodated with a light mediator which is weakly coupled to Standard Model particles.   We discuss a model where the large cross-section generates both the observed rate and is consistent with the results of null experiments.

The outline of this paper is as follows.  In the next section we lay out the rates for standard spin-independent scattering and for the anapole and magnetic dipole operators.  We then turn to discussing the effect of the experimental uncertainties on standard spin-independent, anapole and magnetic dipole scattering cases, and show how, properly accounting for these uncertainties, we can bring the two results into better agreement.  We focus in particular on the sodium quenching factor, xenon prompt photon ($S_1$) to nuclear recoil conversion factor ${\cal L}_{\rm eff}$, the stochasticity of photo-electrons in XENON10, and the systematic uncertainty in the CDMS-Si energy threshold.  Lastly, we discuss models that generate the observed event rates, and conclude.

\section{Scattering Rates}

We begin by reviewing the standard scattering rates and then turn to a discussion of the anapole and dipole rates.  The rate for scattering is
\beq
\frac{dR}{dE_R} = N_T \frac{\rho_\chi }{m_{\chi}} \int_{|\vec{v}| > v_{min}} d^3v v f(\vec{v},\vec{v}_e)\frac{d\sigma}{d E_R},
\label{totalrate}
\eeq
where
\beq
v_{min}=\frac{\sqrt{2 m_N E_R}}{2\mu_N},
\eeq
and $\mu_N$ is the reduced mass of the nucleus-dark matter system.
We take the velocity distribution $f(\vec{v},\vec{v}_e)$ to be a modified Boltzmann distribution:
\beq
f(\vec{v},\vec{v}_e) \propto \left(e^{-(\vec{v}+\vec{v}_e)^2/v_0^2}-e^{-v_{esc}^2/v_0^2}\right) \Theta(v_{esc}^2-(\vec{v}+\vec{v}_e)^2).
\eeq
The additional term is to allow for a smooth cut-off of the velocity distribution near the galactic escape velocity $v_{esc}$.
The Earth's speed relative to the galactic halo is $v_e = v_\odot + v_{orb} \cos \gamma \cos[\omega(t-t_0)]$ with $v_\odot = v_0 + 12 \mbox{ km/s}$, $v_{orb} = 30 \mbox{ km/s}$, $\cos \gamma = 0.51$, $t_0 = \mbox{ June 2nd}$ and $\omega = 2 \pi/\mbox{year}$.  We take as a standard case $v_0 = 220 \mbox{ km/s}$ but allow $v_0$ to vary up to 270 km/s, with higher values decreasing the tension
with XENON10.  We fix $v_{esc} = 500 \mbox{ km/s}$; allowing this to vary changes the DAMA and CoGeNT windows only by a small amount, though the constraints from XENON10 become more stringent for larger $v_{\rm esc}$.

A standard calculation relates the differential rate for scattering off nuclei to the scattering rate off a nucleus $\sigma_N$,
\beq
\frac{d\sigma}{d E_R} = \frac{m_N \sigma_N}{2 \mu_N^2 v^2}.
\label{eq:ratenorm}
\eeq
For the standard spin-independent case, this rate is related to a scattering off protons, $\sigma_p$, through
\beq
\sigma_N = \sigma_p \frac{\mu_N^2}{\mu_n^2} \frac{\left[f_p Z + f_n(A-Z)\right]^2}{f_p^2} F^2(E_R),
\eeq
where $\mu_n$ is the DM-nucleon reduced mass and $f_p$ and $f_n$ are the dark matter couplings to the neutron and proton.
We set $f_n = 0$ since this is the normalization that will arise most naturally
in the models we discuss later,\footnote{Choosing $f_p=f_n$ would give slightly different
results. In particular, $A^2/Z^2$ is approximately 30\% larger for germanium
than for silicon, and thus the CDMS-Si constraints on CoGeNT would be
weakened by approximately this amount.  Similarly, the region of parameter space
favored by sodium scattering
at DAMA would move up about 20\% in cross-section relative to CoGeNT,
which would help (hurt) agreement between the two regions in the case
of scattering through the dipole or anapole (standard) coupling.} and we consider two different choices for the form factor $F^2(E_R)$, which can give rise to ${\cal O}(20\%)$ variations in the derived cross-sections.  We make use of a Helm form factor
\beq
F(E_R) = \frac{3 j_1(q r_0)}{(q r_0)} e^{-(qs)^2 {\rm fm}^2 /2},
\eeq
with two different choices for $r_0$:
\beq
r_0 = \left((1.2 A^{1/3})^2-5 s^2\right)^{1/2} \mbox{ fm}
\label{r01}
\eeq
with $s = 1$,
and
\beq
r_0 = \left((1.23 A^{1/3}-a)^2+\frac{7}{3} \pi^2 b^2 - 5 s^2 \right)^{1/2} \mbox{ fm},
\label{r02}
\eeq
with $a = 0.6$, $b = 0.52$ and $s = 0.9$ \cite{gondoloFF}.

We next derive the rate for scattering through the anapole operator, Eq.~(\ref{anapole}).  The photon coupling to nuclei is
\beq
{\cal O}_N = A^\mu \bar{N}(p)\left(F_1(q) (p+p')_\mu + (F_1(q)+F_2(q)) 2 i\Sigma_{\mu \nu} q^\nu \right) N(p').
\label{eq:photonnucleuscoupling}
\eeq
where $F_1(q),~F_2(q)$ are form factors, and the spin tensor
$\Sigma_{\mu \nu}$ is a generator in the appropriate representation of the Lorentz group for
spin-$J$ nuclei $N$.
For instance, for spin-1/2 nuclei, $\Sigma_{\mu\nu} = \frac{1}{2} \sigma_{\mu\nu}$,
and for spin-0
nuclei $\Sigma_{\mu \nu} = 0$.
In (\ref{eq:photonnucleuscoupling}), the fields $N$ have the standard non-relativistic normalization, which
for spin-1/2 nuclei differs from the standard relativistic normalization
by a factor of $\sqrt{2m_N}$.  The form factors satisfy $F_1(0) = Z,
(F_1(0)+F_2(0)) = \frac{1}{2J}\frac{m_N}{m_p}\frac{b_N}{b_n}$, where $b_N$ denotes
the nuclear magnetic moment and $b_n=e/2m_p$ denotes the Bohr magneton,
since we are already using the more common symbols $\mu_N,\mu_n$ for
reduced masses.\footnote{We are assuming here that the coupling to the nucleus goes through the photon.  For a more general coupling through a dark force only, the magnetic moment and charge can be allowed to float, shifting our results.}  In the non-relativistic limit, the nuclear magnetic moment coupling
can be written $b_N \frac{\vec{J}}{J} \cdot \vec{B}$. We take the $q$-dependence of $F_1(q)$
from the Helm form factor, and we neglect the $q$-dependence of $F_2(0)$.
Making these substitutions, the resulting matrix element, for a Dirac state, is
\beq
\frac{1}{4}\sum |{\cal M}|^2 = \frac{4 m_\chi^2 m_N^2 }{M^4}\left(4 v^2 Z^2F(E_R)^2 - q^2\left(\frac{(m_\chi+m_N)^2}{m_\chi^2 m_N^2} Z^2 F(E_R)^2 - 2 A^2 \frac{J+1}{3J} \frac{b_N^2}{m_N^2 b_n^2} \right)\right).
\eeq
The resultant scattering cross-section, the analogue of Eq.~(\ref{anapolexsec}) and
which should be inserted in Eq. \ref{eq:ratenorm} to obtain the
differential rate, is
\beq
\sigma_N = \frac{\mu_N^2}{4\pi M^4} \left(4 v^2 Z^2 F(E_R)^2 - q^2\left(\frac{(m_\chi+m_N)^2}{m_\chi^2 m_N^2} Z^2 F(E_R)^2 - 2 A^2 \frac{J+1}{3J} \frac{b_N^2}{m_N^2 b_n^2} \right)\right).
\label{anapolecrosssection}
\eeq
When reporting cross-sections for the anapole case, we use a convention closely related to Eq.~(\ref{anapolecrosssection}), taking $\tilde{\sigma} = \mu_n^2/4\pi M^4$.

Similarly, the rate through the magnetic moment operator, Eq.~(\ref{dipole}), can be computed.  We find the resultant scattering cross-section is
\beq
\sigma_N = \frac{4 \mu_N^2}{\pi M^4 \Lambda^2} \left(4 q^2 v^2 Z^2 F(E_R)^2 - q^4\left(\left(\frac{2}{m_N m_\chi}+\frac{1}{m_N^2}\right) Z^2 F(E_R)^2 - 2 A^2 \frac{J+1}{3J} \frac{b_N^2}{m_N^2 b_n^2} \right)\right).
\label{dipolecrosssection}
\eeq
When reporting cross-sections, we use the convention $\tilde{\sigma} = 4\mu_n^2/\pi M^4$.

In the appendix we offer analytic expressions for the velocity integrals in Eq.~(\ref{totalrate}) necessary for computing the total rates in both the standard case and in the case of $v^2$ dependence in the rate.  We next discuss our results using these expressions for the anapole operator with experimental uncertainties folded in.

\section{Results}

Neither DAMA nor CoGeNT measures the total nuclear recoil energy $E_R$.  Instead, both experiments measure the ``electron equivalent'' energy, $E_{ee}$, which is energy deposited into electrons measured as scintillation or ionization.  To extract the nuclear recoil energy, and hence the energy spectrum of the dark matter recoils, one must fold in the quenching factor $Q$ which relates the two.  These quenching factors can be highly uncertain.  In germanium, relevant for CoGeNT, we take two possible parameterizations relating the nuclear recoil and electron-equivalent energies
\beq
Q_{Ge} = 0.19935 A_Q \left(\frac{E_R}{\mbox{ keV}}\right)^{0.1204} ,
\label{GeQ1}
\eeq
and
\beq
Q_{Ge} =  0.224 A_Q \left(\frac{E_R}{\mbox{ keV}}\right)^{0.1204} E^{-\sqrt{0.00383 \mbox{ keV}/E_R}} .
\label{GeQ2}
\eeq
$A_Q$ is an amplitude for the quenching factor that is allowed to vary between 0.85 and 1 in accordance with the uncertainty in the observed quenching factors.

For DAMA's sodium iodine, the quenching factors carry larger uncertainties.  Though DAMA reports a quenching factor $Q_{Na} = 0.30\pm 0.01$ averaged over 6.5 to 97 keV nuclear recoil, other measurements seem to suggest that the measured error on the quenching factor is much larger.  For example, \cite{quench1} reports a sodium quenching factor $Q_{Na} = 0.4\pm 0.2$ over 5-100 keV.  Other experiments report smaller quenching factors with smaller errors: $0.275\pm 0.018$ over 4-252 keV \cite{quench2}, and $0.25 \pm 0.06$ at 10 keV \cite{quench3}.  On the other hand, in many cases it is anticipated that the quenching factor will rise at the lower recoil energies associated with DAMA, as shown in \cite{tretyak}.  We will take the standard value $Q_{Na} = 0.3$, but show the effect of allowing it to rise as high as 0.45 at 5 keV.

\begin{figure}[t!]
\begin{center}
\includegraphics[width=0.45\textwidth]{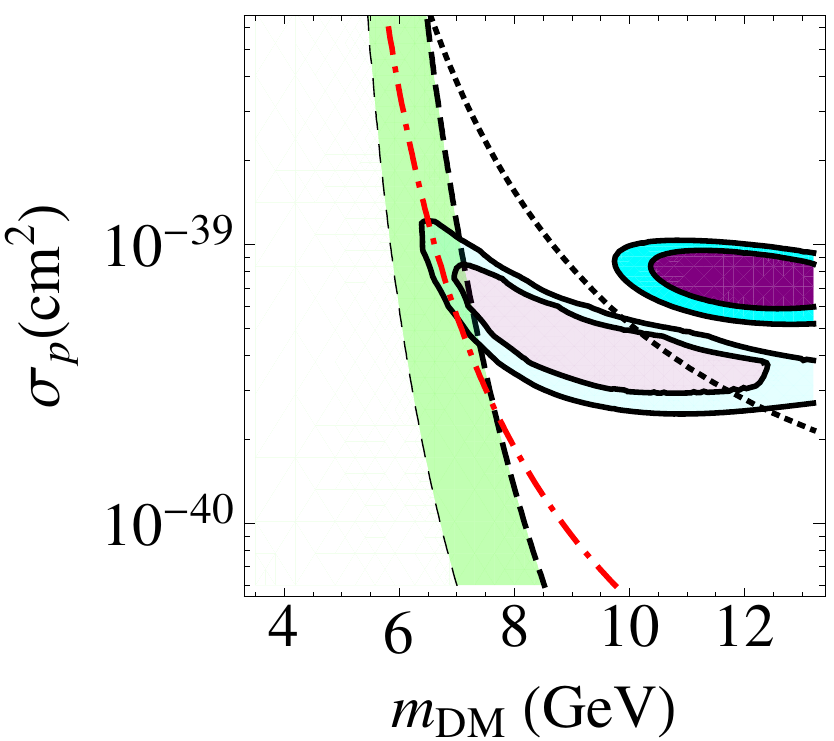}
\includegraphics[width=0.45\textwidth]{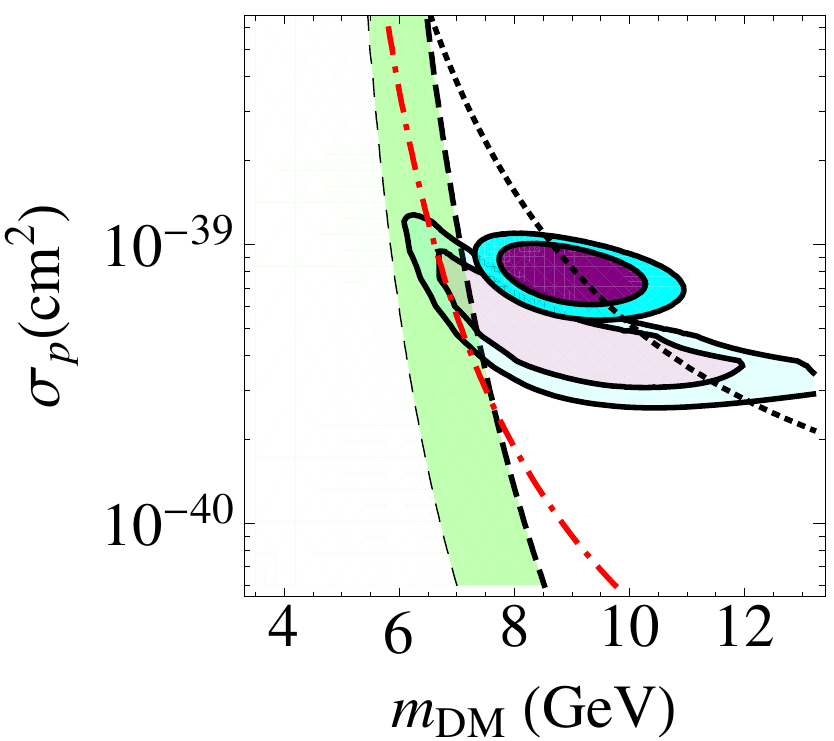}

\caption{{\em left panel:} Allowed regions (90 and 99\% C.L., corresponding to purple and blue) for standard spin-independent scattering, $Q_{Na} = 0.3$, $Q_{Ge}$ from Eq.~(\ref{GeQ1}).  DAMA regions are shown in a darker color than the CoGeNT regions.   A green band shows 90\% exclusion regions from XENON10 depending on the extrapolation of ${\cal L}_{\rm eff}$ below threshold (central values of \cite{Manzur} are taken and extrapolated to remain constant (light dashed) below threshold, or to drop linearly to zero (dark dashed); these extrapolations correspond roughly to Case 1 and Case 2 of \cite{sorensen}).  CDMS-Si (red dot-dashed) and SIMPLE (short dashed) constraints are also shown.
{\em right panel:} Same as left panel, but with $Q_{Na} = 0.45$ and $Q_{Ge}$ from Eq.~(\ref{GeQ2}).
\label{fig:standard1}}
\end{center}
\end{figure}

We first show in Fig.~(\ref{fig:standard1}) the results for the standard case, with the quenching factors $Q_{Na} = 0.3$ and $Q_I = 0.09$ and the Germanium quenching factor parametrized by Eq.~(\ref{GeQ1}).  We take $v_0 = 220 \mbox{ km/s}$, $v_{esc} = 500 \mbox{ km/s}$ and the standard Helm form factor with the standard $r_0$ given by Eq.~(\ref{r01}).  One sees that poor agreement is obtained between CoGeNT and unchanneled DAMA\footnote{We focus only on unchanneled DAMA in this discussion, following \cite{gondolounchanneled}.}.  One can also see that poor agreement between DAMA and the null results of CDMS-Si is obtained.  For the CDMS-Si curves, we take the constraints derived from the set-up of \cite{CDMSIISi,CDMSIISi2}, and we derive results from the SIMPLE experiment for 14 kg-days \cite{SIMPLE}.

Obtaining constraints from XENON10 is more involved.  Recently, it was pointed
out that the effects of stochasticity in the number of $S_1$ photo-electrons
(PEs)
observed can be important for constraining DM in the low mass window
\cite{XENON100}.  In particular, each nuclear recoil energy can be mapped onto
the number of prompt $S_1$ PEs expected in each scatter.
To suppress background contamination, XENON10 cuts out events with fewer
than two PEs, which for a given ${\cal L}_{\rm eff}$
roughly corresponds to a recoil energy threshold $E_{\rm thr}$ given
implicitly by $E_{\rm thr} = 0.37 / {\cal L}_{\rm eff}(E_R)$.
However, there can be low-energy recoil events below $E_{\rm thr}$ that,
thanks to Poisson fluctuations and detector resolution, nevertheless
make it over the S1 cut.  Since the DM signal is highly peaked near
low energies, this results in more DM signal events being tipped above
threshold by the fluctuations than below, resulting in a tighter constraint.
 This was utilized by XENON100 to purportedly rule out the CoGeNT window \cite{XENON100}.  In that analysis, however, a high ${\cal L}_{\rm eff}$ converting expected $S_1$ photo-electrons to nuclear recoil energy was employed.  A choice of a larger ${\cal L}_{\rm eff}$ translates to a lower energy threshold and a strong constraint on light DM.  We think it likely that the ${\cal L}_{\rm eff}$ chosen there is too high, and hence the constraints are too strong. Here we focus on XENON10 constraints, which are in general more restrictive than those due to XENON100 on account of its higher threshold.

We will follow the procedure outlined in \cite{sorensen}, and make use of
the detector resolution and efficiency derived in that analysis.  In particular,
we show in each constraint plot the 90\% limit based on two different
assumptions on ${\cal L}_{\rm eff}$. Both assumptions take the central values
measured for ${\cal L}_{\rm eff}$ in \cite{Manzur}, but the first
(stronger) constraint assumes a constant ${\cal L}_{\rm eff}$ below the
lowest measured energies, and the second (weaker) constraint assumes
a linear interpolation to zero at vanishing recoil energy.
 The constraint on a WIMP model is determined by the predicted rate of
events after convolving with the detector resolution and taking into
account detector and cut efficiencies:
\beq
\frac{dR}{dE_{obs}} = \int_{2 keV} d E_R \frac{dR}{dE_R}
\eta(E_R) \frac{dN}{dE_{obs}}(E_R)\epsilon(E_{obs}).
\eeq
Here, $\eta$ and $\epsilon$ are detector and cut efficiencies (including the acceptance rate of the ``50\%'' acceptance box), taken
from \cite{sorensen} and \cite{XENONinelastic} respectively, and
$dN/dE_{obs}$ is the detector resolution also from \cite{sorensen},
which is somewhat broader than a pure Poisson distribution as utilized in \cite{savagestoch}.
We further cut off the low end of the integration over recoil energies
at $E_R = 2 \mbox{ keV}$, though due to the low detection efficiency at such low
energies, the constraint is not strongly dependent on the exact limit.  The constraints derived from these two extrapolations of ${\cal L}_{\rm eff}$ are shown as the edges of the green band in the figures.

\begin{figure}[t!]
\begin{center}
\includegraphics[width=0.45\textwidth]{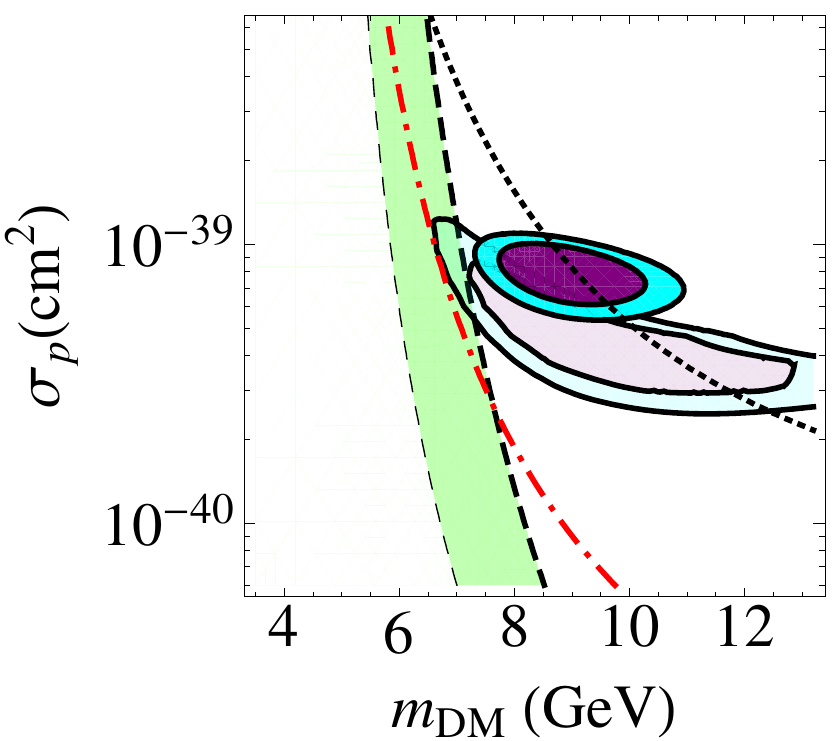}
\includegraphics[width=0.45\textwidth]{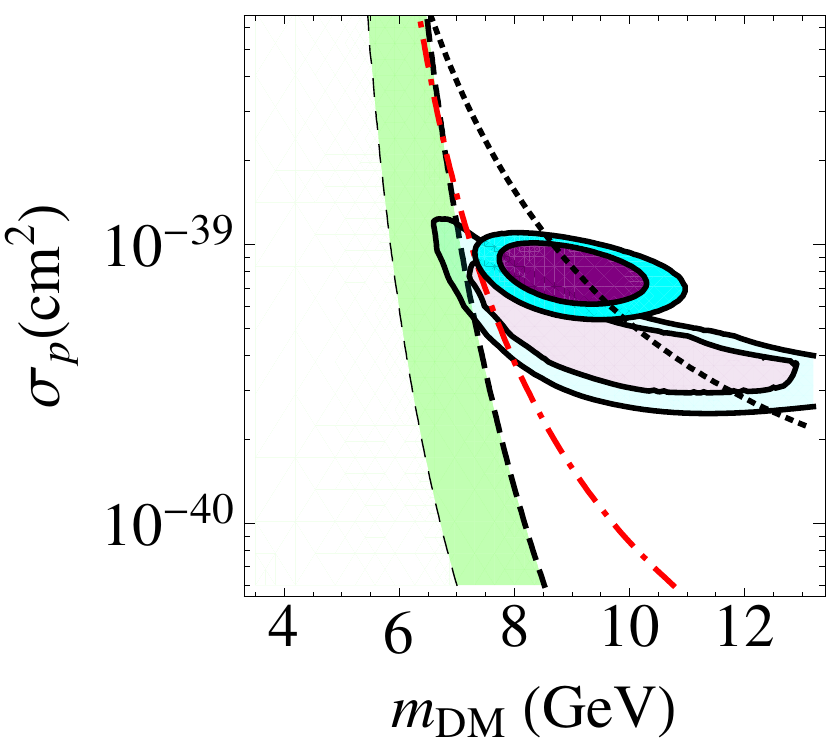}

\caption{{\em left panel:} Allowed regions (90 and 99\% C.L., corresponding to purple and blue) for standard spin-independent scattering, $Q_{Na} = 0.45$, $Q_{Ge}$ from Eq.~(\ref{GeQ2}) with $A_Q = 0.85$.  DAMA regions are shown in a darker color than the CoGeNT regions.   A green band shows 90\% exclusion regions from XENON10 depending on the extrapolation of ${\cal L}_{\rm eff}$ below threshold (central values of \cite{Manzur} are taken and extrapolated to remain constant (light dashed) below threshold, or to drop linearly to zero (dark dashed)).  CDMS-Si (red dot-dashed) and SIMPLE (short dashed) constraints are also shown.
{\em right panel:}  Same as left panel, but with form factor from Eq.~(\ref{r02}) and 20\% threshold uncertainty in CDMS-Si taken into account.
\label{fig:standard2}}
\end{center}
\end{figure}

Agreement between DAMA and CoGeNT can be improved by choosing a different set of experimental parameters.  In the second panel of Fig.~(1), we show the results with $Q_{Na} = 0.45$ and $Q_I = 0.09$ and the Germanium quenching factor parametrized by Eq.~(\ref{GeQ2}).   Some improvement can be obtained by shifting the quenching factor amplitude in Eq.~(\ref{GeQ2}) $A_Q$ to 0.85 as shown in Fig.~(\ref{fig:standard2}), left panel.  We also modify the form factor $r_0$ from Eq.~(\ref{r02}), and take into account a possible 20\% uncertainty in the energy threshold of CDMS-Si \cite{jeter}.  This is shown in Fig.~(\ref{fig:standard2}), right panel.   However, even with this extreme set of parameters, only marginal agreement can be found between the two results.

\begin{figure}[t!]
\begin{center}
\includegraphics[width=0.45\textwidth]{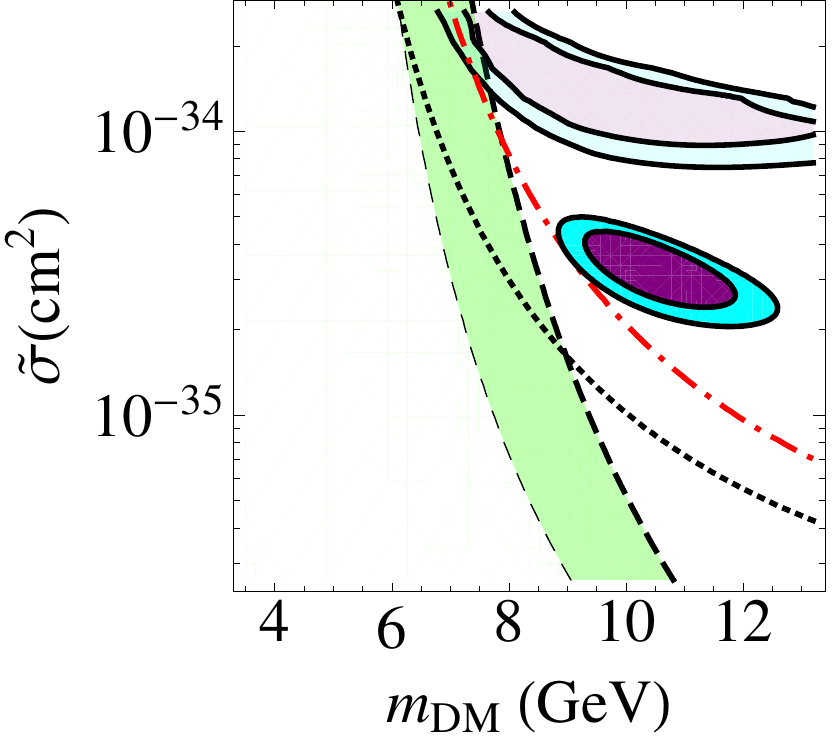}
\includegraphics[width=0.45\textwidth]{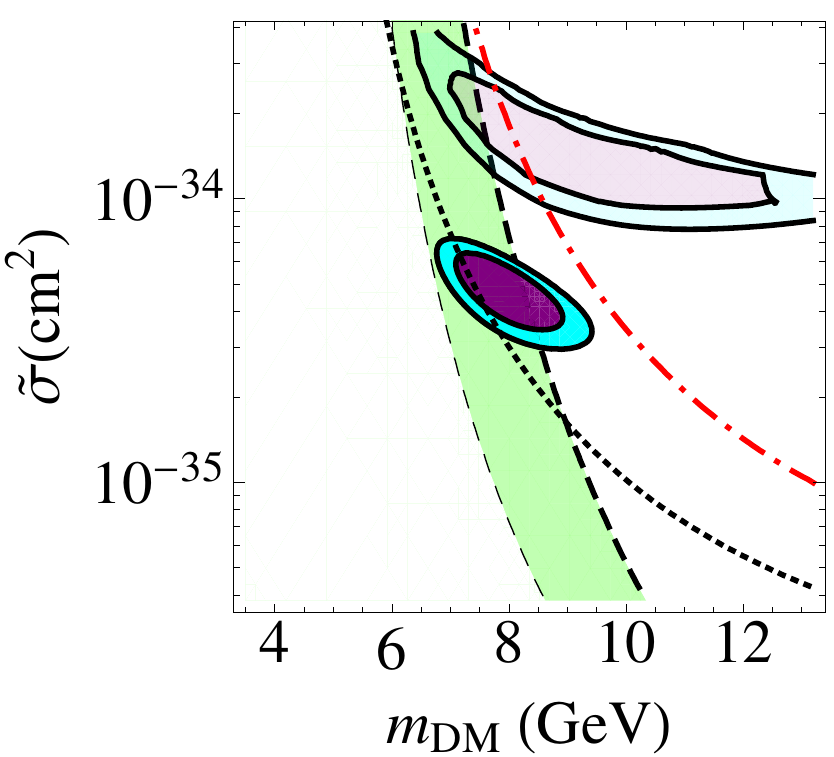}

\caption{{\em left panel:} Allowed regions (90 and 99\% C.L.) for scattering through the anapole operator, $Q_{Na} = 0.3$, $Q_{Ge}$ from Eq.~(\ref{GeQ1}) with $A_Q=1$, and form factor from Eq.~(\ref{r01}).   DAMA regions are shown in a darker color than the CoGeNT regions.   A green band shows 90\% exclusion regions from XENON10 depending on the extrapolation of ${\cal L}_{\rm eff}$ below threshold (central values of \cite{Manzur} are taken and extrapolated to remain constant (light dashed) below threshold, or to drop linearly to zero (dark dashed)).  CDMS-Si (red dot-dashed) and SIMPLE (short dashed) constraints are also shown.
{\em right panel:}  Same as left panel, but with $Q_{Na} = 0.45$ and 20\% threshold uncertainty in CDMS-Si taken into account.
\label{fig:anapole1}}
\end{center}
\end{figure}

We next consider velocity dependent cross-sections, and we look at the anapole operator first.  We show results first for $Q_{Na} = 0.3$ and the quenching factor Eq.~(\ref{GeQ2}) in Fig.~\ref{fig:anapole1}.  We can see that the CoGeNT region moves up relative to the DAMA region, so much so that it lands {\em above} the DAMA region.  Agreement can be improved by taking $Q_{Na} = 0.45$ as shown in Fig.~(\ref{fig:anapole1})b.  For the cases that the operator pushes the CoGeNT region above the DAMA region, we take $A_Q = 1$ and $r_0$ from Eq.~(\ref{r01}) since both of these choices tend to push CoGeNT down somewhat relative to DAMA (in comparison to $A_Q = 0.85$ and $r_0$ from Eq.~(\ref{r02})).  While the agreement between the two regions is poor, agreement of either experiment with the constraints from XENON10 and CDMS-Si is improved, especially for DAMA.  The results of experiments sensitive to large spin-dependent cross-sections, such as COUPP \cite{COUPP} and PICASSO \cite{PICASSO}, may also be relevant since both the anapole and magnetic dipole scattering cross-sections have contributions from a spin-dependent interaction.  We find, however, that neither of these experiments significantly constrain the otherwise allowed parameter space, so we do not show these results on our plots.

\begin{figure}[t!]
\begin{center}
\includegraphics[width=0.45\textwidth]{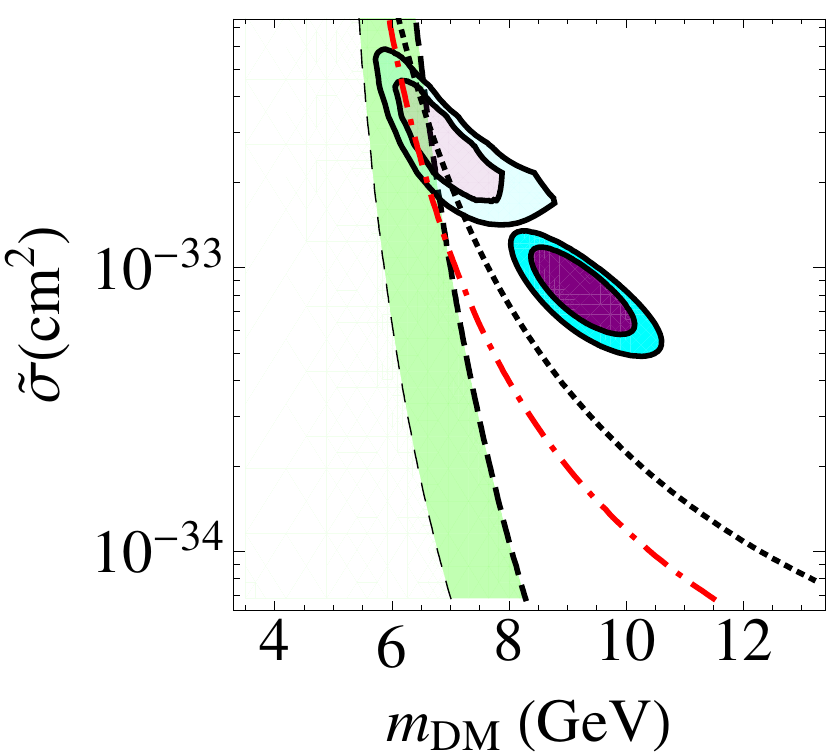}
\includegraphics[width=0.45\textwidth]{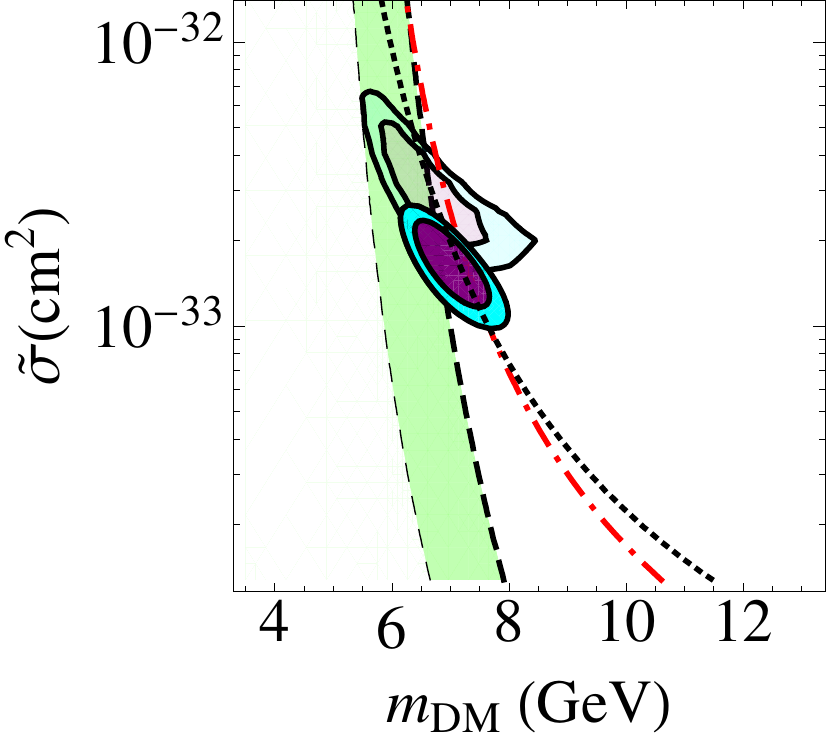}
\caption{{\em left panel:} Allowed regions (90 and 99\% C.L.) for scattering through the magnetic dipole operator, $Q_{Na} = 0.3$, $Q_{Ge}$ from Eq.~(\ref{GeQ1}) with $A_Q=1$, and form factor from Eq.~(\ref{r01}).   DAMA regions are shown in a darker color than the CoGeNT regions.   A green band shows 90\% exclusion regions from XENON10 depending on the extrapolation of ${\cal L}_{\rm eff}$ below threshold (central values of \cite{Manzur} are taken and extrapolated to remain constant (light dashed) below threshold, or to drop linearly to zero (dark dashed)).  CDMS-Si (red dot-dashed) and SIMPLE (short dashed) constraints are also shown.
{\em right panel:}  Same as left panel, but with $Q_{Na} = 0.45$ and 20\% threshold uncertainty in CDMS-Si taken into account. \label{fig:dipole}}
\end{center}
\end{figure}

\begin{figure}[t!]
\begin{center}
\includegraphics[width=0.45\textwidth]{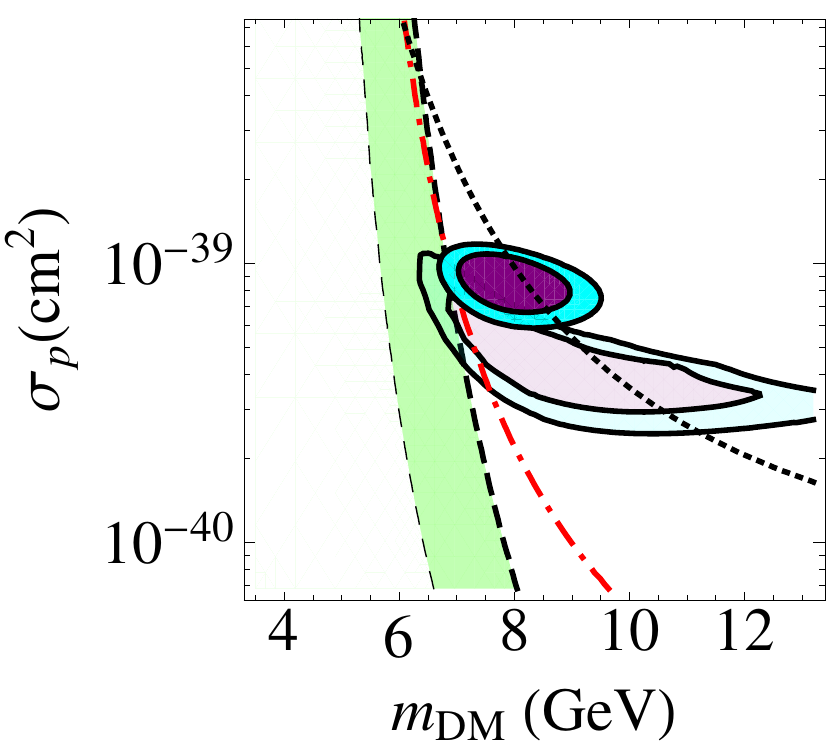}
\includegraphics[width=0.45\textwidth]{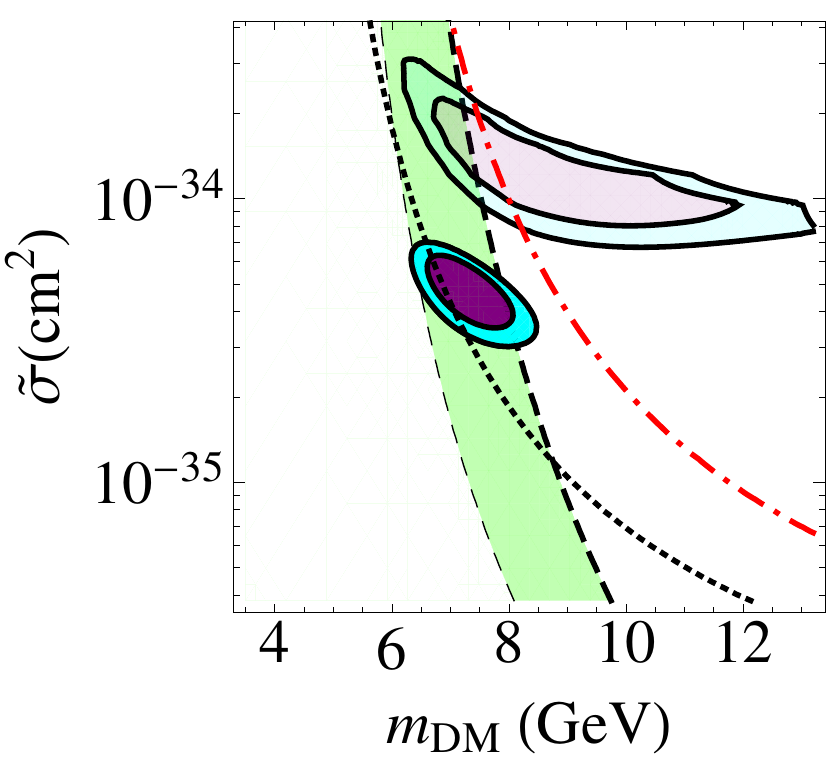}
\caption{ {\em left panel:} Same as Fig.~(\ref{fig:standard2})b (standard
WIMP coupling), but with $v_0 = 270 \mbox{ km/s}$.
{\em right panel:} Same as Fig.~(\ref{fig:anapole1})b (scattering through anapole
operator), but with $v_0 = 270 \mbox{ km/s}$.
\label{fig:270}}
\end{center}
\end{figure}

\begin{figure}[t!]
\begin{center}
\includegraphics[width=0.45\textwidth]{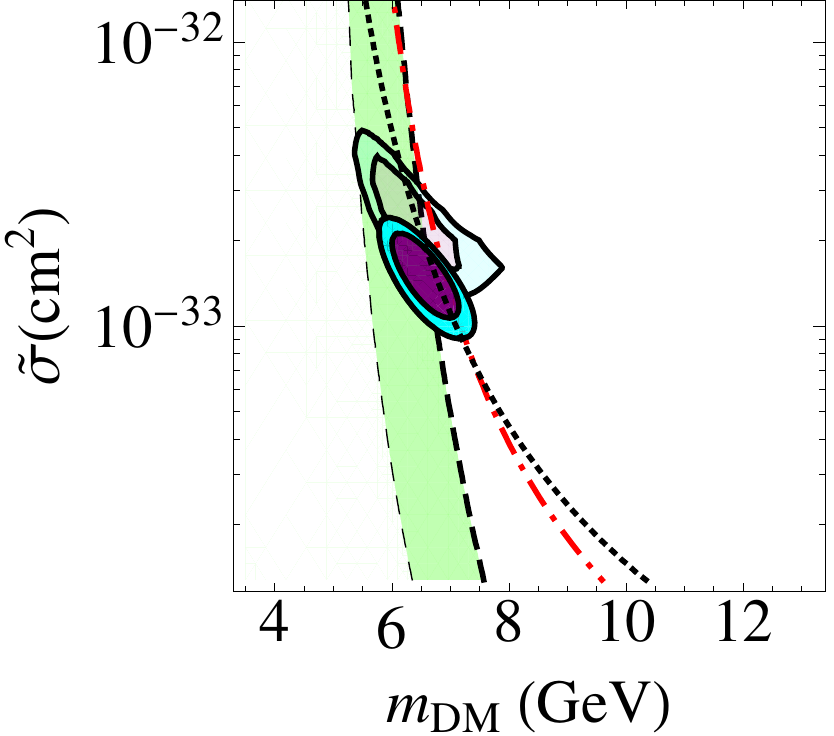}
\caption{ Same as Fig.~(\ref{fig:dipole})b (scattering through magnetic dipole operator), but with $v_0 = 270 \mbox{ km/s}$.  One can see that optimal agreement of DAMA and CoGeNT with each other and with the results of the null experiments is obtained for scattering through this operator with this set of astrophysical and experimental parameters (most importantly, $Q_{Na} = 0.45$ here).
\label{fig:dipole270}}
\end{center}
\end{figure}

We consider in Fig.~(\ref{fig:dipole}) the effect of the magnetic dipole operator on the scattering regions of DAMA and CoGeNT, taking $\Lambda = 100 \mbox{ MeV}$.  We can see that of all the operators optimal agreement between DAMA and CoGeNT is obtained, and improved agreement with the null results of XENON10 can be obtained.
In addition, we note the effect of changing $v_0$ to 270 km/s is to gain marginal improvement over $v_0 = 220 \mbox{ km/s}$ in terms of the compatibility of the regions with each other and with the results of the null experiments as shown in Figs.~(\ref{fig:270}),~(\ref{fig:dipole270}).

\begin{figure}[t!]
\begin{center}
\includegraphics[width=0.473\textwidth]{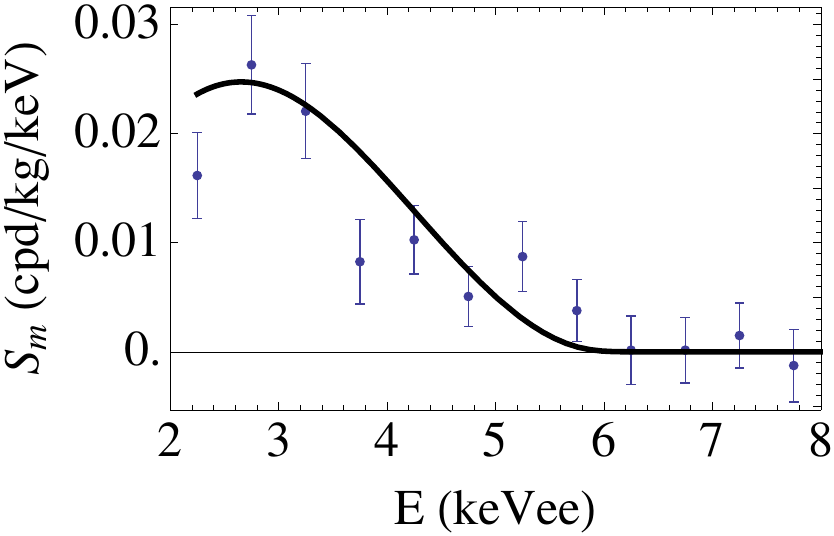}
\includegraphics[width=0.45\textwidth]{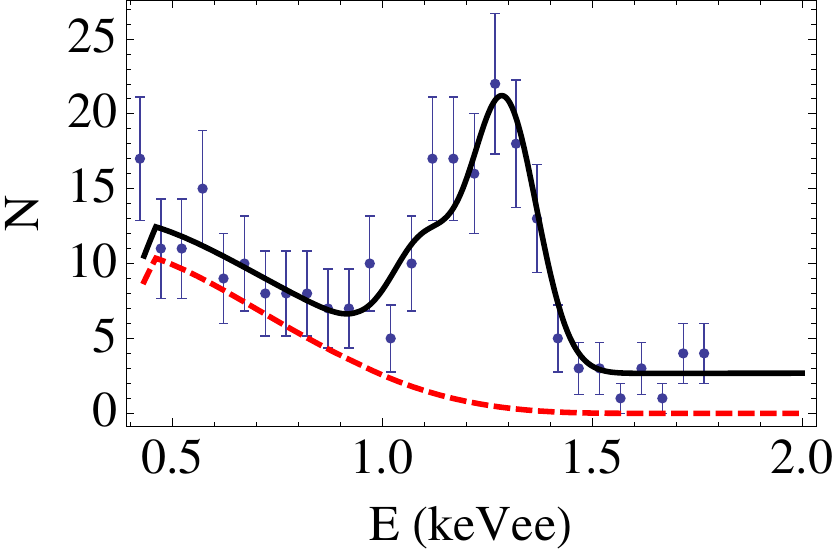}
\caption{ The spectra observed at DAMA (left) and CoGeNT (right) for a benchmark magnetic dipole scattering with $m_\chi = 6.4 \mbox{ GeV}$,~$\tilde{\sigma}= 1.9\times 10^{-33}\mbox{ cm}^2$.  The rate is corrected for CoGeNT efficiencies, and these efficiencies are the reason for the drop in the rate in the lowest bin; the uncorrected rate continues to rise there. In the right panel, the black solid line is the total predicted event rate including signal and background, whereas the red dashed line shows just the signal rate.
\label{fig:spectra}}
\end{center}
\end{figure}

Since agreement is optimal between DAMA, CoGeNT and the results of the null experiments for the magnetic dipole operator, we show in Fig.~(\ref{fig:spectra}) a sample spectrum generated for the magnetic dipole operator with $m_\chi = 6.4 \mbox{ GeV}$,~$\tilde{\sigma} = 1.9\times 10^{-33}\mbox{ cm}^2$.  The model reproduces very well the DAMA spectrum, but struggles to obtain a large enough rise in CoGeNT.  The reason for this is the dramatic drop in efficiencies at the low recoil energy. The uncorrected rate continues to rise.
CoGeNT additionally has data on events in one extra bin below their energy
threshold, with 26 events, which they discount. This must either be explained
by a large boundary effect or large errors in the efficiency near
threshold. Neglecting this, both fits are quite good; $\Delta \chi^2 \equiv \chi^2 - \chi^2_{\rm min}$
is 2.7 for both CoGeNT and DAMA (90\% agreement is $\Delta \chi^2 = 4.91$).
$\chi_{\rm min}^2$ is 21.78 for 17 bins for DAMA and 15.16 for 27 bins
for CoGeNT.  It also remains to be explored whether such models could be responsible for the 32 events reported by CRESST-II in their oxygen band \cite{CRESSTII}, though it is difficult to compare concrete models with no report of the total exposure or efficiency after cuts.

We have shown here than when experimental uncertainties are taken into account appropriately, that a region of parameter space where DAMA and CoGeNT are consistent with the results of null experiments is available.  Agreement can be improved by considering the effects of the scattering primarily through the magnetic dipole operator.  We now turn to a brief discussion of models where the anapole and magnetic dipole operators are the dominant form of scattering.

\section{Models}

Though different in detail, both the anapole and magnetic dipole operators
are velocity and momentum suppressed, and thus need sufficiently large cross-sections to explain the
event rates seen at DAMA and CoGeNT.  Therefore the mass of the
dark photon $A_\mu$ that mediates the interaction should be fairly light.
For example, consider the mass of the mediator necessary to generate the large cross-sections for scattering through the anapole interaction, Eq.~(\ref{anapole}).  That cross-section scales as
\beq
\tilde{\sigma} = \frac{\mu_n^2}{4 \pi M^4} = 10^{-34} \mbox{ cm}^2 \left(\frac{44 \mbox{ GeV}}{M}\right)^4,
\eeq
where $10^{-34}$ cm$^2$ is approximately the size needed for $\tilde{\sigma}$, from
Fig. \ref{fig:anapole1}.  A cross-section of this size is not difficult
to generate.  Consider a model with a Weyl fermions $\chi$
and complex scalar $\phi$ with charges $+1, -2$, respectively.
A second Weyl fermion $\chi^c$ is present for anomaly cancellation, but
otherwise plays no role; we impose a $Z_2$ symmetry under which
$\chi$ is odd and all other fields are even in order to restrict the
possible interactions of $\chi^c$ with $\chi$.  The allowed renormalizable
interactions within the dark sector are then the following:
\begin{eqnarray}
{\cal L} &\supset& \bar{\chi} \sigma^\mu D_\mu \chi
+ \bar{\chi}^c \sigma^\mu D_\mu \chi^c
 + | D_\mu \phi|^2 + V(|\phi|^2)
+ \lambda \phi \chi \chi + \lambda' \phi^* \chi^c \chi^c  + h.c.
\end{eqnarray}
We assume
that $\phi$ dynamically obtains a vev $v = \langle \phi \rangle
\sim 10 $ GeV, which gives mass to $\chi$ and the dark gauge boson.
The $\chi$ mass term after symmetry-breaking is Majorana, which gives
a simple explanation for why the anapole operator dominates:
the leading vector operator $\bar{\chi} \gamma^\mu \chi \bar{N} \gamma_\mu N$
vanishes for Majorana fermions.  The dark Majorana particles can couple to
the dark force because the dark gauge group is broken.
Other interactions of $\chi$ with the
dark force are higher dimensional and therefore suppressed relative
to the anapole interaction. Parity is badly broken in the dark sector, and
a dark electric dipole moment (EDM), while higher-dimensional, has the
same $q^2$
suppression as the anapole interaction.  However, such a dark EDM must
be generated radiatively and therefore be phase-space suppressed relative
to the anapole.

The dark sector then interacts with the Standard Model through kinetic mixing
$\epsilon$ of the light dark force with field strength $f_{\mu\nu}$ with
hypercharge:
\beq
{\cal L}\supset -\frac{1}{4} f_{\mu\nu}f^{\mu\nu} + \epsilon f_{\mu\nu} B^{\mu\nu} .
\eeq
Then we find
\beq
\tilde{\sigma} = 10^{-34} \textrm{cm}^2
\left( \frac{\epsilon}{2 \times 10^{-3}} \right)^2
\left( \frac{100 \textrm{MeV}}{m_M} \right)^2
\left( \frac{8 \textrm{GeV}}{v} \right)^2 ,
\eeq
where $m_M = \sqrt{2} g_D v$ is the gauged messenger mass and $g_D$ is the
dark gauge coupling.  These choices for $\epsilon$ and $m_M$ are
consistent with the bounds on kinetic mixing.

The magnetic dipole operator may also be easily generated with
a sufficiently large cross-section.
It arises quite naturally when the DM is a Dirac fermion composite.  Consider the case
where the dark matter $\chi$ is a fermionic bound state with compositeness
scale $\Lambda$.  If $\chi$ is not charged under the dark gauge
force, but its constituents are (similar to the neutron and
the electromagnetic force), then at low momentum transfer its
interactions with the Standard Model will shut off.  We require that the scale
$\Lambda$ be above the momentum transfer $\sim 30 $ MeV relevant
for scattering at DAMA, so that  we may parametrize the interactions
of $\chi$ with the dark gauge field in terms of the lowest
dimensional gauge-invariant operator, $\bar{\chi} \sigma^{\mu\nu} \chi
F_{\mu\nu} /\Lambda$.
Here we are assuming that parity is not violated in the dark sector,
so that a dark EDM cannot be generated.  In our constraint plots, we
have already taken $\Lambda = 100$ MeV, since this is approximately the minimum
that the compositeness scale can be and still give a reliable effective
theory.  With this choice, we find
\beq
\tilde{\sigma} = \frac{4 \mu_n^2}{\pi m_M^4} = 1.5  \times 10^{-32.} \textrm{cm}^2
\left(\frac{\epsilon}{2\times 10^{-3}} \right)^2 \left( \frac{ 600 \textrm{MeV}}
{m_M} \right)^4 .
\eeq

The additional $q^2$ dependence in the cross-section that arises
from the dark magnetic dipole moment tends to lower the direct detection
rates at low energies.  This allows lighter DM, since the main obstacle to
taking dark matter to be very light is the rapidly rising spectrum
at low energies in DAMA and CoGeNT, ruining the fit at low recoils.
The $q^2$ dependence tends to counteract this, allowing for successful models which evade the constraints of XENON10, even taking the stochasticity into account.

\section{Summary}

We have studied the implications of experimental uncertainties and non-standard velocity dependence on the agreement of CoGeNT and DAMA both with each other and with the results of the null experiments.  While some marginal agreement can be obtained between the two experiments when the sodium quenching factor is pushed to $Q_{Na} = 0.45$ and the CDMS-Si energy threshold is assumed to have a systematic error of 20\% (assumed to be too low), optimal achievement between the two regions is not obtained for the standard spin-independent case.  Agreement can be improved by choosing a different velocity dependence, and in particular the magnetic dipole operator gives optimal agreement with all experiments.
Simple models were constructed where velocity and momentum dependent cross-sections are expected to dominate, and it was shown that acceptably large cross-sections can be obtained.

While the DAMA and CoGeNT signals can be consistent with each other and the results of null experiments, large theoretical and experimental uncertainties limit our current understanding of the signals and their consistency with models.  As we are learning, in an age of DM discovery, we must systematically quantify errors on theoretical and experimental parameters to determine whether a given DM model is consistent with the signals and with the results of null experiments.  Until these uncertainties are reduced and the low mass signals can be strongly excluded given the uncertainties, it appears the light DM candidate cannot be ruled out.  We look forward to further pursuing theoretically well-motivated models of low mass DM.

\section*{Acknowledgments}

We thank Katie Freese, Dan Hooper, Brian Feldstein, Aaron Pierce,  and Chris Savage for discussions, and Peter Sorensen for help on interpreting the XENON stochasticity bound.  We also thank the Aspen Center for Physics for hospitality while this work was being completed.
ALF is supported by DOE grant DE-FG02-01ER-40676 and NSF CAREER grant PHY-0645456.

{\em Note added:}  While this work was nearing completion, Refs.~\cite{ChangMag,Barger} appeared which also explore the magnetic dipole operator.
Additionally, after this work was submitted to arXiv, we became aware of a talk \cite{peterstalk} discussing preliminary results from a new method of analyzing XENON10 data that would allow it to probe lower recoil energies.  Once finalized, this method will likely have implications for the scenarios discussed in this paper.

\section{Appendix}

Noting that $\vec{v} \cdot \vec{v}_e = v v_e \cos \chi$ and $d^3 v = 2 \pi v^2 dv d\cos\chi$, one finds for the standard velocity integral in Eq.~(\ref{totalrate}):
\beq
\int_{|\vec{v}| > v_{min}}  \frac{f(\vec{v},\vec{v}_e)}{v}d^3v = \frac{1}{N} \alpha_{SI} \Theta(x_{esc} -(x_e + x_{min})) +\frac{1}{N} \beta_{SI} \Theta(x_{esc}-x_{min}+x_e,-x_{esc}+x_{min}+x_e),
\label{standardrate}
\eeq
where
\begin{eqnarray}
\alpha_{SI} = && \sqrt{\pi} \left(\mbox{erf}(x_{min}+x_e) - \mbox{erf}(x_{min}-x_e)\right)-4 x_e e^{-x_{esc}^2}\left(1+x_{esc}^2-\frac{x_e^2}{3}-x_{min}^3)\right), \\
\beta_{SI} = &&\sqrt{\pi}\left(\mbox{erf}(x_{esc}) -\mbox{erf}(x_{min}-x_e) \right) \\
           &-& 2 e^{-x_{esc}^2}\left(x_{esc}+x_e-x_{min}-\frac{1}{3}(x_e-2 x_{esc}-x_{min})(x_{esc}+x_e-x_{min})^2\right),
\end{eqnarray}
with
\beq
N =4 x_e v_0 \left[\frac{1}{2} \sqrt{\pi }
   \text{ erf}(x_{esc})-e^{-x_{esc}^2} x_{esc} \left(\frac{2
   x_{esc}^3}{3}+1\right)\right].
\eeq

Likewise one can compute the relevant integral in Eq.~(\ref{totalrate}) for $v^2$ dependent cross-sections.  It is
\beq
\int_{|\vec{v}| > v_{min}}  v f(\vec{v},\vec{v}_e)d^3v = \frac{1}{N} \alpha_{v^2} \Theta(x_{esc} -(x_e + x_{min})) +\frac{1}{N} \beta_{v^2} \Theta(x_{esc}-x_{min}+x_e,-x_{esc}+x_{min}+x_e),
\label{vsq}
\eeq
where
\begin{eqnarray}
\alpha_{v^2} & = & -\frac{1}{6} e^{-(x_{min}+x_e )^2} \left(3 \sqrt{\pi } \left(2 x_e
   ^2+1\right) \text{erf}(x_{min}-x_e )
   e^{(x_{min}+x_e )^2} \right. \nonumber \\
   && \left. -3 \sqrt{\pi } \left(2 x_e
   ^2+1\right) \text{erf}(x_{min}+x_e )
   e^{(x_{min}+x_e )^2}  -6 x_e  e^{4
   x_{min} x_e }+24 x_{esc}^2 x_e  e^{-x_{esc}^2+(x_{min}+x_e )^2} \right. \nonumber \\
   && \left. -6
   x_{min} e^{4 x_{min} x_e }+6
   x_{min}-6 x_e +8 x_e ^3 e^{-x_{esc}^2+(x_{min}+x_e
   )^2}+24 x_e  e^{-x_{esc}^2+(x_{min}+x_e )^2}\right) \nonumber \\
  &&-e^{-x_{esc}^2}\left( -\frac{2x_e^5}{15}+\frac{4 x_e^3 x_{esc}^2}{3}-2 x_e
   \left(x_{min}^4-x_{esc}^4\right)\right)
\end{eqnarray}

\begin{eqnarray}
\beta_{v^2} & = & 2 e^{-x_{esc}^2} \left[\frac{1}{3} \left(x_{min}^3-(x_{esc}+x_e )^3\right)-\frac{1}{4}
   e^{-x_{min}^2-x_e ^2 + x_{esc}^2} \left(\sqrt{\pi }
   \left(2 x_e ^2+1\right) e^{x_{min}^2+x_e ^2}
   (\text{erf}(x_{min}-x_e ) \right. \right. \nonumber \\
   &&  \left. -\text{erf}(x_{esc})) -2 e^{2
   x_{min} x_e } (x_{min}+x_e )+2 (x_{esc}+2 x_e )
   e^{x_{min}^2+x_e ^2-x_{esc}^2}\right) - \frac{1}{30} \left(-x_{e}^5+10 x_{e}^3 x_{esc}^2 \right.  \nonumber \\
   && \left. \left. +10
   x_{e}^2 \left(2 x_{esc}^3+x_{min}^3\right)+15 x_{e}
   \left(x_{esc}^4-x_{min}^4\right)+4 x_{esc}^5-10
   x_{esc}^2 x_{min}^3+6 x_{min}^5\right) \right]
\end{eqnarray}

The standard velocity integral Eq.~(\ref{standardrate}) (weighted with $q^2$ or $q^4$) can be combined with the $v^2$ integral to give the total rate for the scattering through the anapole and magnetic dipole operators.

\end{document}